\documentstyle[prl,epsfig,aps] {revtex}
\begin{document}
\def\ll{\label}
\def\re{\ref}
\def\c{\cite}
\def\b{\begin}
\def\La{\Lambda}
\def\r1{(\ref{$1})}
\def\ot{\otimes}
\def\nn{\nonumber}
\def\sn{\rm sn}
\def\pa{\partial}
\def\kap{\kappa}
\def\ms{\medskip}
\def\cR{\cal R}
\def\cF{\cal F}
\def\cP{\cal P}
\def\ti{\tilde}
\def\cn{\rm cn}
\def\dn{\rm dn}
\def\ga{\gamma}
\def\ep{\epsilon}
\def\th{\theta}
\def\ba{\begin{array}{c}}
\def\e{\end}
\def\sk{\smallskip}
\def\ea{\end{array}}
\def\pr{\prod}
\def\ni{\noindent}
\def\si{\sigma}
\def\da{\dagger}
\def\De{\Delta}
\def\de{\delta}
\def\bet{\beta}
\def\ov{\over}
\def\ha{{1\over 2}}
\def\qr{{1\over 4}}
\def\l{\left}
\def\l({\left(}
\def\r){\right)}
\def\r{\right}
\def\rw{\rightarrow}
\def\om{\omega}
\def\la{\lambda}
\def\al{\alpha}
\def\sec{\section}
\def\be{\begin{equation}}
\def\bc{\begin{center}}
\def\ec{\end{center}}
\def\bit{\begin{itemize}}
\def\eit{\end{itemize}}
\def\ee{\end{equation}}
\def\ed{\end{document}}
\def\bea{\begin{eqnarray}}
\def\eea{\end{eqnarray}}
\def\efr{\end{flushright}}
\def\nn{\nonumber\\}

\title{Coupled nonlinear Schr\"odinger equations with cubic-quintic nonlinearity:
Integrability and soliton interaction in non-Kerr media }
\author{R. Radhakrishnan $^{1}$, A. Kundu $^{2}$ and M. Lakshmanan$^{1}$ }
\address{$^{1}$ Centre for Nonlinear Dynamics, Department of Physics, Bharathidasan
University, Tiruchirapalli 620 024, India.\\
$^{2}$ Theory Group. Saha Institute of Nuclear Physics, 1/AF, Bidhan Nagar,
Calcutta  700 064, India. 
}
\maketitle 
\begin{abstract}
We propose an integrable system of  coupled nonlinear Schr\"odinger
equations with cubic-quintic terms describing the effects of quintic
nonlinearity on the ultra-short optical soliton pulse propagation in
non-Kerr media.  Lax pair, conserved quantities and
 exact soliton solutions for the proposed integrable model are
given. Explicit form of two-solitons  are used to study soliton interaction
showing many intriguing features including inelastic (shape changing) scattering.
Another novel system of coupled equations with fifth-degree nonlinearity is
derived, which represents vector generalization
of the known chiral-soliton bearing  system.
\end{abstract}
\pacs{PACS number(s): 42.81Dp, 42.65Tg, 03.40kf}
\vskip1pc 

\section {INTRODUCTION}
Optical solitons have promising potential to  become principal carriers
 in telecommunication due to their capability of propagating long distances
without attenuation and changing their shapes \cite{1,2,3,4}. Therefore
considerable attention is being paid theoretically and experimentally to
analyse the dynamics of optical solitons in optical waveguides (for example
silica fibers) under different contexts \cite{1,2,3,4,5,6}.  Such
investigations are helpful for realizing  optical soliton
applications, particularly in  soliton-based optical communication
systems \cite{5} and nonlinear optical switches \cite{6}.  The waveguides
used in such optical systems are usually of Kerr type \cite{7}. 
Consequently the dynamics of light pulses are described by the nonlinear
Schr\"odinger (NLS) family of equations with cubic nonlinear terms
\cite{7,8}. However, as the intensity of the incident light field becomes
stronger, non-Kerr nonlinearity effect comes into play and due to this
additional effect, the physical features and the stability of NLS soliton
can change \cite{3}.

The way through which non-Kerr nonlinearity influences NLS soliton propagation
is described by the NLS family of equations with higher-degree nonlinear terms
\cite{9,10,11,12,13,14,15,16}.  Therefore  investigations on these evolution
equations become important from a theoretical point of view. Particularly
this importance has received a boost after the experimental observation of
multistability of solitons in non-Kerr fiber \cite{17}.  In general the
models proposed in the literature \cite{3,9,10,11,12,13,14,15,16} for
describing the non-Kerr
 effects are not completely integrable and can not be solved exactly by the
inverse scattering transform method. In such
non-integrable systems, therefore, the details of 
 soliton-interaction  during collision can not be described exactly 
 and hence are still open to debate.  However, numerical
stimulations \cite{18} show that even the slightest change from the Kerr
nonlinearity results in the two solitons annihilating each other, merging or
creating many new solitons, depending on the initial inclination of the two
solitons and their shapes.  But besides the important problem of computer
time, the numerical approach is not very appealing in the sense that it is
not a simple task to get physical insight from purely numerical experiments. 
The idea therefore is  to use approximate analytical methods such as perturbation
technique, variational method, etc., in order to compensate for the lack of
exact results \cite{19}.  By treating the quintic nonlinear terms due to
non-Kerr effect as perturbations of the cubic NLS equations, i.e.
restricting  the effects of
quintic nonlinearity to be less predominant than the  cubic terms, the 
NLS equations are studied both
analytically and numerically in \cite{3,10,13,16}.

In this paper we have obtained for the first time (to our knowledge) an
integrable system of coupled NLS equations including cubic-quintic terms describing the
effects of quintic nonlinearity with arbitrary coupling, which generalizes 
  the coupled hybrid NLS equations
with cubic nonlinearity \cite{20,21}. Lax pair as well as infinite set of conserved quantities are
derived for the proposed integrable model.  We also find exact soliton
solutions for our model and using the explicit form of the two-soliton
solution we study the associated soliton collision. A remarkable interrelation
between the newly proposed integrable model and the celebrated Manakov model
\cite{22} helps to use the recently discovered general two-soliton solution
of the later model \cite{23}. This reveals  
the  fascinating occurrence of shape changing inelastic 
 soliton collision also in the 
present model, in addition to some other interesting features. 
We believe that such a study using higher-order solitons
(multisolitons) becomes
important in the light of the  proposal 
by Hasegawa and Nyu \cite{24} regarding "eigenvalue communication",
 in which the information may be transmitted
by higher-order solitons and a recent one sugested 
by Jakubowski, Steiglitz and Squiev \cite{25}
 on nontrivial information transmission system, which uses the more general
two-soliton solution of the Manakov system \cite{22} derived by
Radhakrishnan, Lakshmanan and Hietarinta in \cite{23}.
We also find that the Hamiltonian  of the present integrable model is
associated with a non-canonical Poisson bracket (PB) structure. However,
using the same Hamiltonian with  canonical PB relations we
derive  another coupled system with quintic nonlinearirity, which may
be transformed to a vector generalization of  the chiral-solitonic model
of   Aglietti et al \c{jakiw}.

The plan of our paper is as follows. The basic evolution equation with
cubic-quintic nonlinearity which describes the soliton evolution in non-Kerr
media with parabolic nonlinearity is discussed in Sec-II.  Sections III 
is devoted to the proposed model and presents its integrability property
by explicit construction of the Lax pair as well as the hierarchy of
conserved quantities generated through a recurrence relation, which in turn
is derived from a coupled Ricatti equation using the Lax operator. Sec  IV gives the
 exact soliton solution   for our cubic-quintic nonlinear evolution equation
and using the explicit analytic solution we study here the collision
process of solitons so as to understand the influence of quintic nonlinear
effects on the Manakov model. Section V establishes the 
interrelation and gauge transformation between our system and the 
Manakov model  including the anyon-like nonultralocal 
PB  for our model.  Sec VI presents a vector generalization of the chiral
soliton bearing system starting from the same Hamiltonian and discusses
some of  its
 interesting features. Sec VII  gives  concluding remarks.

\section {BASIC EVOLUTION EQUATIONS}

Generally when high optical intensities (or materials with high nonlinear
coefficients even at moderate optical intensities, for example semiconductor
doped glasses, organic polymers, thin liquid filled capillaries, etc.) are
considered, it is  necessary to take into account higher power
nonlinearities arising from an expansion of the refractive index in powers
of intensity $ I $ of the light pulse : $ n = n_0 + n_2 I + n_4 I^2 + ..., $
where $ n_0 $ is the linear refractive index coefficient and $ n_2, n_4..., $
are nonlinear refractive index coefficients \cite{3,10}.

In the case of $ n = n_0 + n_2 I + n_4 I^2 $, the wave equation for high-intensity
light pulse propagation in an isotropic single-mode optical fiber with a
circular cross-section and fiber axis z can be written as
\begin{equation}
{\nabla}^2 {\bf E} - { 1 \over c^2} { {\partial}^2 {\bf D_L} \over
{\partial {t}^2} } = {1 \over c^2} { {\partial}^2 {\bf D_{NL}}\over
{\partial {t}^2} },
\end{equation}
where $ c $ is the speed of light, the linear part $ {\bf D_L} $ and the nonlinear
part $ {\bf D_{NL}} $ of the electric field displacements are related to the
electric field $ {\bf E}({\bf r},t) $ by the relation $ {\bf D_L} = $
$ \int_{0}^{\infty} { \epsilon (t') {\bf E} (t-t')dt' } $ and $ {\bf D_{NL}}
= \epsilon_2 |E|^2 + \epsilon_4 |E|^4 $, in which $ \epsilon = n_o^2 $,
$ \epsilon_2 = 2 n_2 n_0 $ and $ \epsilon_4 = 2 n_4 n_0 $.

A solution of Eq.(1) is sought in the form
\begin{equation}
{\bf E} = {\bf e} R(r) A(z,t)e^{i\beta z - i \omega t},
\end{equation}
where $ {\bf e} $ is a unit vector in the direction of wave polarisation,
$ R({\bf r}) $ describes the transverse field modes, in which $ {\bf r} $
is a two dimensional vector in the x-y plane and $ A(z,t) $ is a slowly varying
amplitude.  Here we assume that $ R({\bf r}) $, which is
mainly defined by the linear
effects, corresponds to the modal distribution of the fundamental fiber mode
$ HE_{11} $, for simplicity. Then from the equations (1) and (2), assuming
the temporal dispersion of the dielectric permitivity to be small, using
the slowly varying envelope approximation and following the procedure in
section 2.1 of \cite{8}, the follwing nonlinear partial differential equation
for $ A(z,t) $ can be obtained,

\begin{eqnarray}
i \left[ A_z + { 1 \over v_g} A_t \right] - { 1 \over 2} k_{\omega\omega}
A_{tt}
- { i \over 6} k_{\omega\omega\omega} A_{ttt}
 + { kn_2 \over n_0 } \alpha_0 | A |^2 A \\ \nonumber
+ { kn_4 \over n_0} \beta_0 |A|^4 A +
 { i n_2 \alpha_0 \over v_gn_0}
(|A|^2 A)_t +  { i n_4 \beta_0 \over v_gn_0}(|A|^4 A)_t = 0,
\end{eqnarray}
where the subscript $ \omega$ of the wave number $ k $ ( i.e. $k_\omega $) indicates differentiation
of k with respect to $ \omega$, the subscripts $ t $ and $ z $ of $ A $ indicate
differentiation of $ A $ with respect to the coordinates $ t $ and $ z $
respectively and the numerical values of the parameters $ \alpha_0 $ and
$ \beta_0 $ depend on the form of the function $ R({\bf r}) $.

It is convenient now to transform the above equation to a reference frame moving with
group velocity $ v_g $, and to introduce dimensionless variables $ q = { A \over
|A_0| } $, $ \gamma = { {2n_4 \beta_0 |A_0|^2} \over {n_2 \alpha_0} } $,
$ \gamma_1 = { -k_{\omega \omega \omega } \over {3 (- k_{ \omega \omega }) } }
\left[ {1 \over { z_{NL}(-k_{\omega \omega})} } \right]^{1 \over 2}$,
$ \gamma_2 = {2 \over v_g} \left[ { {n_2 \alpha_0 |A_0|^2} \over { k n_0
(-k_{ \omega \omega} )} } \right ]^{1 \over 2}$, 
$ \gamma_3 = {2n_4 \beta_0 |A_0|^2} \over {v_g} \left[ { {|A_0|^2} \over { k n_0 n_2 \alpha_0
(-k_{ \omega \omega} )} } \right ]^{1 \over 2}$,
$ z_{NL}^{-1} = { {k n_2 \alpha_0 |A_0|^2} \over n_0} $,
$ t \rightarrow \left[ {1 \over {z_{NL} (-k_{ \omega \omega} )} } \right]^ {1
\over 2 } ( t - {z \over v_g} ) $ and $ z \rightarrow {z \over 2 z_{NL} } $,
in which $ z_{NL} $ characterizes the nonlinear propeties of the fiber
and $ |A_0| $ is a measure of the maximum amplitude of the input pulse.
Now equation (3) takes the form
\begin{equation}
iq_z +  q_{tt} + 2 |q|^2 q + \gamma |q|^4 q + i \gamma_1
q_{ttt} + i \gamma_2 (|q|^2q)_t + i \gamma_3(|q|^4 q)_t = 0.
\end{equation}
Equation
 (4) describes the effects of quintic nonlinear terms proportional
to the real parameters $ \gamma $ and $ \gamma_3 $ on the dynamics of pulse
envelope allowing self-phase modulation and higher-order linear and nonlinear
dispersions.  For  pulse widths greater than 100 fs, one can neglect the last
three terms of (4) and the resulting equation is a well-studied
\cite{3,10} simple normalized NLS equation with cubic-quintic
nonlinear terms.  Further the system (4) is a special case of the dynamical
equation considered for the fiber system with saturating nonlinearity
\cite{12}.

It is of further interest to extend the above analysis to include multimode
effects.  For this purpose, there are several ways to generalize
 equation (4) to a set of coupled
equations depending on the physical situations. A
fairly general form of coupled nonlinear Schr\"odinger (CNLS) equation with
cubic-quintic nonlinearity is

\begin{eqnarray}
iq_{1z}+q_{1tt}+2(|q_1|^2+B|q_2|^2)q_1+ \gamma(|q_1|^2+B
|q_2|^2)^2q_1+\rho q_1 + \kappa q_2 &&
\nonumber \\
-i \mu [\gamma_1 q_{1ttt} +
\gamma_2 [ ( |q_1|^2 + B|q_2|^2 ) q_{1t}
+ ( q_1 q_{1t}^{*}+ B q_2 q_{2t}^{*} ) q_1 &&
\nonumber \\ 
+ \gamma^{\prime}
( q_{1t} q_{1}^{*}+ B q_{2t} q_{2}^{*} ) q_1 ] 
+ \gamma_3 [  ( |q_1|^2 + B|q_2|^2 )^2 q_1 ]_t
]  = 0. && \nonumber  \\
iq_{2z}+q_{2tt}+2(B|q_1|^2+|q_2|^2)q_2+ \gamma(B|q_1|^2+
|q_2|^2)^2q_2-\rho q_2 + \kappa q_1 &&
\nonumber \\
-i \mu [\gamma_1 q_{2ttt} +
\gamma_2 [ ( B|q_1|^2 + |q_2|^2 ) q_{2t}
+ ( B q_1 q_{1t}^{*}+  q_2 q_{2t}^{*} ) q_2 &&
\nonumber \\ 
+ \gamma^{\prime}
( B q_{1t} q_{1}^{*}+ q_{2t} q_{2}^{*} ) q_2 ] 
+ \gamma_3 [  (B |q_1|^2 + |q_2|^2 )^2 q_2 ]_t]  = 0. 
&& \label{5}
\end{eqnarray}
A nonlinear direction coupler with quintic nonlinearity(or parabolic nonlinearity
coupler) has $ B=\rho=\mu= 0 $ \cite{26}.  For $ \mu=\gamma=0 $, Eq.(5) acts
as a mathematical model for a periodically twisted elliptical birefringent
fiber \cite{27}. If $ \gamma = \rho = \kappa = \gamma_1 = \gamma_3 = 0 $ ,
and $ \gamma^{\prime}$ =1  then Eq.(3)
becomes the coupled hybrid nonlinear Schr\"odinger equation \cite{20} used
to investigate the effects of birefringence on pulse propagation in the
femto-second range. In the absence of quintic nonlinear terms proportional to the
real parameters $ \gamma $ and $ \gamma_3 $, soliton interaction supported
by  system (5) has been studied by deriving higher-order soliton solutions
under the parameteric
restrictions  $ B=1 $ and $ 3 \gamma_1 = \gamma_2 $ \cite{28}.  One may
also note that when $ B=1 $ and $\gamma' = 1$ the linear coupling terms proportional to the
parameter  $ \rho $ and $ \kappa $ in Eq.(5) can be removed without affecting the other
terms by using the transformation
\begin{eqnarray}
q_1 \rightarrow cos( {\theta \over 2} ) e^{i \Gamma z} q_1 -
sin ( {\theta \over 2}) e^{-i \Gamma z} q_2, \nonumber \\
q_2 \rightarrow sin ( {\theta \over 2} ) e^{i \Gamma z} q_1 +
cos ( {\theta \over 2}) e^{-i \Gamma z} q_2, 
\end{eqnarray}
where $ \Gamma  = ( \rho^2 + \kappa^2 ) ^{1 \over 2} $ and $ \theta
= tan^{-1}({\kappa \over \rho}). $
If the nonlinearity is restricted only to cubic terms corresponding to pulse
widths greater than 100fs, one obtains 
 the  celebrated  integrable  Manakov
model \cite{22,33} 
\begin{eqnarray}
iq_{1Mz}+q_{1Mtt}+2(|q_{1M}|^2+|q_{2M}|^2)q_{1M} &=&0,  \nonumber
\\
iq_{2Mz}+q_{2Mtt}+2(|q_{1M}|^2+|q_{2M}|^2)q_{2M} &=&0.
\ll{manakov} \end{eqnarray}
There is a large amount of theoretical work 
\cite{1,2,3,4,5} devoted to  the CNLS
family of equations with cubic nonlinearity.  However, to the knowledge of
the authors, to date the CNLS equations with non-Kerr nonlinearity have
received very little attention in the literature, particularly in connection
with the integrability aspects.  In the following sections, by identifying one
such integrable nonlinear evolution equation, we derive the two-soliton
solution so as to get some idea about soliton interaction in non-Kerr media.

\section{Integrability property of the proposed model: 
Lax pair and conserved quantities}
It is evident that eq. (5) does not exhibit the explicit rotational symmetry
in the internal space spanned by the vector $  (q_1, q_2). $ However,
for  $B=1 $ such a symmetry is restored. Assuming further that
 $\gamma_1 = \gamma_3 =\gamma^{\prime}=0, $ (5) can be reduced to the 
following  quintic generalization of the  coupled  cubic NLS equation,
\begin{eqnarray}
iq_{1z}+q_{1tt}+2(|q_1|^2+|q_2|^2)q_1+ \gamma (|q_1|^2+|q_2|^2)^2q_1
+ \rho q_1 + \kappa q_2 -i \mu \gamma_2 
&& \nonumber \\
\left[ (|q_1|^2+|q_2|^2)q_{1t} +
(q_1 q_{1t}^{*} + q_2 q_{2t}^{*})q_1 \right] = 0, && \nonumber \\
iq_{2z}+q_{2tt}+2(|q_1|^2+|q_2|^2)q_1+ \gamma (|q_1|^2+|q_2|^2)^2q_2
- \rho q_2 + \kappa q_1 -i \mu \gamma_2 
&& \nonumber \\
\left[ (|q_1|^2+|q_2|^2)q_{2t} +
(q_1 q_{1t}^{*} + q_2 q_{2t}^{*})q_2 \right] = 0.
&& 
\ll{eqn}\end{eqnarray}
 The terms $
\rho $ and $ \kappa $ can be removed   from (\re{eqn}) 
 by using  transformation (6). Eq. (\re{eqn}) without quintic nonlinearity
was investigated in \c{21}.However the remarkable fact is that (\re{eqn})
itself can be shown to be exactly integrable.
Our proposed model   is  a further generalization of (\re{eqn}) and
naturally of 
(\re{manakov}), where
the internal rotational symmetry is broken again and more parameters 
are introduced with  arbitrary values, which   can be chosen conveniently  
to suit the real situations. The model can be given as
\begin{eqnarray}
iq_{1z}+q_{1tt}+2(|q_1|^2+|q_2|^2)q_1+(\rho _1|q_1|^2+\rho
_2|q_2|^2)^2q_1+2\rho _2[(\tau _1-\rho _1)|q_1|^2+(\tau _2-\rho _2) && 
\nonumber \\
|q_2|^2]\left| q_2\right| ^2q_1-2i[(\rho _1|q_1|^2+\rho
_2|q_2|^2)q_1]_t+2i(\rho _1q_1^{*}q_{1t}+\rho _2q_2^{*}q_{2t})q_1=0, && 
\nonumber \\
iq_{2z}+q_{2tt}+2(|q_1|^2+|q_2|^2)q_2+(\tau _1|q_1|^2+\tau
_2|q_2|^2)^2q_2+2\tau _1[(\rho _1-\tau _1)|q_1|^2+(\rho _2-\tau _2) && 
\nonumber \\
|q_2|^2]\left| q_1\right| ^2q_2-2i[(\tau _1|q_1|^2+\tau
_2|q_2|^2)q_2]_t+2i(\tau _1q_1^{*}q_{1t}+\tau _2q_2^{*}q_{2t})q_2=0.
&& \label{new}
\end{eqnarray}
where $\rho_1, \rho_2, \tau_1 $ and $ \tau_2$ are real free parameters.
It is evident that with  a symmetric reduction  $
\rho_1=\rho_2 = \tau_1 = \tau_2, $  we can recover (\re{eqn}) from (\ref{new}), while
a different  reduction with $q_1=q$ and $q_2=0$ (or $q_1=0$ and $
q_2=q)$ yields the 
integrable  Kundu-Eckhaus equation
\cite{29} \begin{equation}
iq_{z}+q_{tt}+2|q|^2q+ \rho^2_1 |q|^4q
-2i\rho_1 (|q|^2)_t q =0. \label{KE}
\end{equation}
Importantly this generalized model (\re{new}) turns out also to be  exactly
integrable.
For establishing the integrability property of 
the proposed system, which consequently proves also the integrability of the
reduced model
(\re{eqn}), 
we find the Lax pair $( L, M) $ associated with (\re{new}) as
\begin{mathletters}
\begin{eqnarray}
L= \left(
\begin{array}{ccc}
-i\lambda  & q_1                    & q_2 \\
-q_1^*     & \ -i\theta_{1t}+i\lambda & 0   \\
-q_2^*     & 0                      & -i\theta_{2t}+i\lambda  
\end{array} \right), 
\label{11a}\\ 
M= \left(
\begin{array}{ccc}
[-2i\lambda^2+i(|q_1|^2+|q_2|^2)],   & 
2\lambda q_1+iq_{1t}+\theta_{1t}q_1, &
2\lambda q_2+iq_{2t}+\theta_{2t}q_2
\\
-2\lambda q_1^*+iq_{1t}^*-\theta_{1t}q_1^*, &
[2i\lambda^2-i|q_1|^2-i\theta_{1z}],        &
-iq_1^*q_2
\\
-2\lambda q_2^*+iq_{2t}^*-\theta_{2t}q_2^*, &
-iq_1q_2^*,                                 &
[2i\lambda^2-i|q_2|^2-i\theta_{2z}]        
\end{array}
\right),  \label{11b}
\end{eqnarray} 
where
\begin{eqnarray}
\theta _1=\int^t_{-\infty} (\rho
_1|q_1|^2+\rho _2|q_2|^2)dt',\qquad
 \theta _2=\int^t_{-\infty} (\tau _1|q_1|^2+\tau
_2|q_2|^2)dt' . \label{11c}
\end{eqnarray} 
\end{mathletters} 
Here $\lambda $ is the spectral parameter.
 It may be easily checked that    the zero curvature condition 
$ L_z-M_t+[L,M]=0,$ with the explicit Lax operators (11), yields Eq.
(\ref{new}).
In  section V we  give another evidence of its integrability
by relating   system (\ref{new}) to the integrable Manakov model through
a  gauge
transformation of  the 
pair (11)  to the Manakov Lax operators.

Integrable systems, as is well known, possess infinite number of  conserved
quantities in involution, of which usually the lower ones are of physical
importance. Explicit forms of such conserved quantities for the integrable
system (\ref{new}) can be derived from a recurrence relation
obtained from the Ricatti equation related to the Lax operator.
 For this purpose we use the linear system related to  (11a)
\begin{equation} \Phi_t(\lambda,t)=L(\lambda,t) \Phi (\lambda,t),
\ \Phi=(\phi_1,\phi_2,\phi_3)
 \label{linear}
\end{equation} and observe that $$ \ln a(\lambda)= \ln \phi_1 e^{-i\lambda
t}|_{t \rightarrow \infty} =\sum_n
c_n \lambda^{-n} $$ serves as the generator of the  conserved
quantities $\{c_n \}$ through an
expansion in  the spectral parameter $\la $.  The first equation of the system (\ref {linear}) 
 thus yields the relation
\be c_n=
 \int^{+ \infty}_{- \infty} dt (q_1 \Gamma^{(1)}_n
 +q_2 \Gamma^{(2)}_n), \qquad n \geq 1 \ll{cn0} \ee
with the expansion $ \Gamma^{(a)}= \sum_{n=1}^\infty \Gamma^{(a)}_n \lambda^{-n},
a=1,2,$ where we have denoted $ 
\Gamma^{(1)}={\phi_2 \ov \phi_1}$ and $ 
\Gamma^{(2)}={\phi_3 \ov \phi_1}.$ For finding now the infinite set of
conserved quantities  we may use the rest of the
equations of (\re{linear}) to derive 
a set of two coupled Ricatti equations for $\Gamma^{(1)}(\la) $
and  $\Gamma^{(2)}(\la). $ Expanding in powers of $\la $ as mentioned above
we obtain  the recurrent relations
\begin{mathletters} 
\bea -2i\Gamma^{(1)}_{n+1}&=&
 \Gamma^{(1)}_{nt}+ i \th_{1t} \Gamma^{(1)}_{n}+q_1 \sum_{i=1}^{n-1}
\Gamma^{(1)}_{n-i}\Gamma^{(1)}_{i}  +q_2 \sum_{i=1}^{n-1}
\Gamma^{(1)}_{n-i}\Gamma^{(2)}_{i},  
\label{14a}\\
 -2i\Gamma^{(2)}_{n+1}&=&
 \Gamma^{(2)}_{nt}+i \th_{2t} \Gamma^{(2)}_{n}+q_2 \sum_{i=1}^{n-1}
\Gamma^{(2)}_{n-i}\Gamma^{(2)}_{i}  +q_1 \sum_{i=1}^{n-1}
\Gamma^{(2)}_{n-i}\Gamma^{(1)}_{i},  
\label{14b}\eea
\end{mathletters} 
with $\Gamma_1^{(a)}=-\frac{i}{2}q_a^*$. This gives finally the conserved 
quantities in the explicit form as
\bea
c_1 & =& -{1 \ov 2i} \int_{-\infty}^{+\infty} dt (|q_1|^2+|q_2|^2) \ll{c1} \\ 
c_2 & =&- {i \ov 4} \int_{-\infty}^{+\infty} dt [-i (q_1q_{1t}^*+ q_2q_{2t}^*)+ \rho_1|q_1|^4 +
\tau_2|q_2|^4+ (\rho_2+\tau_1)|q_1|^2 
|q_2|^2 ] \ll{c2}\\ 
c_3& =& -{i \ov 8} \int_{-\infty}^{+\infty} dt [(q_1q_{1tt}^*+ q_2q_{2tt}^*)+ (|q_1|^2 +
|q_2|^2)^2+ \nonumber \\
 &i&(|q_1|^2N_{1t}+|q_2|^2N_{2t})+2i
(N_1q_1q_{1t}^*+N_2q_2q_{2t}^*) 
-(N_1^2|q_1|^2+ N_2^2|q_2|^2)  ], 
\ll{c3}\eea
etc., where $N_1=\th_{1t}= \rho_1|q_1|^2+ \rho_2|q_2|^2  $ and 
$ N_2=\th_{2t}=\tau_1|q_1|^2+\tau_2|q_2^2|$.
The above conserved quantities in (\ref{c1}-\re{c3}) may be
interpreted in terms of the number operator $N$, the total momentum $P$ and 
the total energy
or the Hamiltonian of the system $H$. However it is intriguing to remark
that since here the fields $q_a,q_a^*$ do not have canonical Poisson
bracket relations some care has to be taken in deriving the equation of
motion (\ref{new}) from the Hamiltonian (\re{c3}) . In fact the Poisson
bracket structures of the fields, which are derived in sec V, show an
interesting anyon-like feature.

\section{Exact soliton solutions and  scattering of solitons
 in the generalized Manakov model in non-Kerr
media   }
The proposed system (\ref{new}) 
 with quintic nonlinearity  
allows exact $N$-soliton solutions, which can be found for example by
Hirota's method following the same procedure as in the Manakov model.
However a more direct and convenient way is to use the known solutions of
the Manakov model (\ref{manakov})
 themselves for 
constructing the soliton solutions of (\ref{new}). This is possible due to the
interrelation between these two models, which will be established in the
next section.
 Thus we  find the explicit 
1-soliton solution 
of    (\ref{new}) in the form
\be 
(q_{1}, q_{2} ) =
 (\alpha e^{i\de_1 
\tanh (\nu (t-vz+\de))}, \beta e^{i\de_2 
\tanh (\nu (t-vz+\de))}) sech (\nu (t-vz +\de))e^{i(\kappa t+\omega z)}
\ll{s1} ,\ee
where different parameters of the solution are related with the spectral
parameter $\la= \nu +i\kappa$ and the parameters of the model as
$$ v=2 \kappa, \ \  \omega =\nu^2-\kappa ^2, \quad 
 \de_1= {1 \ov \nu}(\rho_1 |\al|^2
+\rho _2|\bet|^2 ), \quad  \de_2= {1 \ov \nu}(\tau_1 |\al|^2
+\tau_2|\bet|^2 ) $$ 
together with a constant phase $\de$. Comparing with the Manakov soliton we see  that
there is an interesting phase change in the carrier wave. The plane wave
like character in the Manakov model has been deformed into a wave suffering
compression and rarification of the phases in a kink-like profile (see Fig.
1a,b).
  This shows that the effect of quintic nonlinearity of our model appears
in the soliton phases and therefore in the derivatives of the soliton
profile $q_{ax}, q_{at},$ which must also change the momentum and energy of
the soliton.

Now exploiting the higher-soliton solutions of the Manakov model it is
also possible to find higher-solitons for (\re{new}) in explicit form.
The Manakov model (\ref{manakov}) has received considerable attention
in recent years \cite{33,34,23} in order to understand the soliton
collision in birerefringent fiber.  However the importance of finding
higher-soliton solutions in explicit form has been understood only
quite recently \cite{23,25}. By constructing the most general
two-soliton solution of the integrable Manakov model,   two of the
authors (R.R and M.L) and Hietarinta\cite{23} have shown that the
soliton in birefringent fiber can in general change its shape after
interaction due to an intensity redistribution among the modes,
eventhough the total intensity remains conserved.  This shape changing
collision arising essentially due to the change in polarisation angle
helps to realise the exciting possibility of switching between
components. (However, the standard shape-preserving collision property
of the (1+1) dimensional soliton system is recoverd, when restrictions
are imposed on some of the free parameters in the two-soliton solution
\cite{23} ). Recently using the shape changing collision concept,
Jakubowski, Steiglitz and Squier \cite{25} have designed sequences of
solitons operating on other sequences of solitons that effect logic
operations and suggested nontrivial information transformation system.

Now in order to investigate the implication  of this property of soliton when
the additional cubic and quintic nonlinear terms are included,
we construct the two-soliton solution of the system (\ref{new}) and
   expect again a non-trivial change in
the soliton phase as in the case of 1-soliton solution.
Such a two-soliton solution expressed compactly  through
that of the 
 Manakov system (\ref{manakov}) $(q_{1M}, q_{2M} ) $ takes the form (see eq.(24) below)
\begin{eqnarray}
q_1 = q_{1M} e^{i \int {(\rho_1 |q_{1M}|^2 + \rho_2 |q_{2M}|^2 ) dt}}, 
\ \ \
q_2 = q_{2M} e^{i \int {(\tau_1 |q_{1M}|^2 + \tau_2 |q_{2M}|^2 ) dt}},
\end{eqnarray}
where a general two-soliton solution\cite{23} of the Manakov model is given by
\begin{equation}
(q_{1M}, q_{2M} ) =
{ { (\alpha_1, \beta_1)e^{\eta_1} + (\alpha_2, \beta_2)e^{\eta_2} +
(e^{\delta_1}, e^{\delta_1^{\prime}}) e^{\eta_1 + {\eta_1}^{*} + {\eta_2} }
+
(e^{\delta_2}, e^{\delta_2^{\prime}}) e^{\eta_1 + {\eta_2} + {\eta_2}^{*} } }
\over
{ 1 + e^{\eta_1 + {\eta_1}^{*} + R_1}  +  e^{\eta_1 + {\eta_2}^{*} + \delta_0}
+ e^{{\eta_1}^* + {\eta_2} + {\delta_0}^*}
+ e^{\eta_2 + {\eta_2}^{*} + R_2} + e^{\eta_1 + {\eta_1}^{*} +
\eta_2 + {\eta_2}^{*} + R_3 } } } ,
\end{equation}
in which $ \eta_j = k_j(t+ ik_j z) $ , $j =1,2 $,
$ e^{\delta_0} = { \kappa_{12} \over {k_1 + k_2^*}} $,
$ e^{R_1} = { \kappa_{11} \over {k_1 + k_1^*}} $,
$ e^{R_2} = { \kappa_{22} \over {k_2 + k_2^*}} $,
$ e^{\delta_1} = { {k_1 - k_2 } \over { (k_1 + k_1^*)(k_1^* + k_2)} }
(\alpha_1\kappa_{21} - \alpha_2 \kappa_{11}) $,
$ e^{\delta_2} = { {k_2 - k_1 } \over { (k_2 + k_2^*)(k_1 + k_2^*)} }
(\alpha_2 \kappa_{12} - \alpha_1 \kappa_{22}) $,
$ e^{\delta_1^{\prime}} = { {k_1 - k_2 } \over { (k_1 + k_1^*)(k_1^* + k_2)} }
(\beta_1 \kappa_{21} - \beta_2 \kappa_{11}) $,
$ e^{\delta_2^{\prime}} = { {k_2 - k_1 } \over { (k_2 + k_2^*)(k_1 + k_2^*)} }
(\beta_2 \kappa_{12} - \beta_1 \kappa_{22}) $,
$ e^{R_3} = { {|k_1 - k_2|^2 } \over { (k_1 + k_1^*)(k_2+ k_2^*)|k_1 + k_2^*|^2 }}
(\kappa_{11} \kappa_{22} - \kappa_{12} \kappa_{21}) $
and $ \kappa_{ij} = {  {(\alpha_i {\alpha_j}^* + \beta_i {\beta_j}^* ) }
\over {k_i + k_j } } $.
The six  arbitrary complex parameters $ \alpha_1 , \alpha_2, \beta_1 ,
\beta_2, k_1 $ and $ k_2 $ determine the amplitude, velocity and phase of
the asymptotic soliton. As we have detected already, we see
also here that the two-soliton solution (19) of  (\ref{new})
differs from that of the Manakov model in   phase-terms in a nontrivial way.
Clearly, the  phase change depends on the values of 
the real free parameters $ \rho_1, \rho_2, \tau_1 $ and $ \tau_2 ,$  and vanishes
for the trivial choice giving back the Manakov soliton.
A natural question therefore arises: Does this change in phase of $ (q_1, q_2) $
 for non-zero values of
$ \rho_1, \rho_2, \tau_1 $ and $ \tau_2 ,$ which in turn accounts for the
effect of the quintic terms in (\re{new}),
make any qualitative change in the
behaviour of soliton collision? This can be directly  studied  using the
two-soliton solution (19-20) parallel to the procedure of the
Manakov model  given in Ref. \cite{23}. In order to answer
the above question,  we have plotted 
pictures of soliton collision corresponding to the function
$ ( |q_{1t}|^2, |q_{2t}|^2 ) $  instead of
$ ( |q_1|^2, |q_2|^2 ), $ since due to   $ ( |q_1|^2, |q_2|^2 ) $ =
$ ( |q_{1M}|^2, |q_{2M}|^2 ) $ one has to look into the derivatives of the
fields, where the effect of phase terms are reflected.
For comparing with the pure Manakov model let us consider first the case
$ \rho_1 = \rho_2 = \tau_1 = \tau_2 = 0. $
 Fig. 2, 
shows the asymptotic forms of $ ( |q_{1t}|^2, |q_{2t}|^2 ) $ for this case 
at $ z = \mp 7 $ for the parameter values $ k_1 = 1.5 + i 0.5 ,
k_2 = 2.0 - i 0.7 $, $ \alpha_1 = \beta_1 = \beta_2 = 1 $, and $ \alpha_2 =
{(39 + i80) \over 89} $.
At $ z=-7 $, we have two well seperated asymptotic profiles as shown in
Fig. 2a.  During the propagation, these two solitary profiles 
interact with each other as shown in Fig. 3 and change
form after interaction. For example, at
$ z = +7 $ they have the profiles as shown in Fig. 2b.  The change
in shape disappears if we apply the elastic collision (shape-preserving
collision) condition, namely $ \alpha_1 : \alpha_2 = \beta_1:\beta_2 $
by following the work of \cite{23}.
In Figs. 2-3, as mentioned, there is a splitting in each of the asymptotic
profiles which appear
before and after interaction. This happens due to  the following reason. In
the case corresponding to the Manakov model 
  one obtains
\begin{equation}
q_{jMt}^{n \mp} \sim A_j^{n \mp} k_{nR} e^{i \eta_{nI}} sech ( \eta_{nR}
+ \phi^{n \mp} ) \left \{ \left[ -k_{nR} tanh( \eta_{nR} + \phi^{n \mp} ) \right]
+ i k_{nI} \right \} |_{z \rightarrow \pm \infty}, j,n=1,2
\end{equation}
where $ \eta_{nR} = k_{nR} (t - 2 k_{nI} z) $, $ \eta_{nI} = k_{nI}t
+ (k_{nR}^2 - k_{nI}^2)z $, the subscript j denotes mode while superscript
$ n \mp $ is used to define the two different interacting solitary waves
appearing at $ z \sim \mp \infty $ and $ A_j ^{n \mp} $ and
$ \phi^{n \mp}  $
determine unit
polarization vector and phase of the modes as defined in Ref. \cite{23}.
From (21)
one can note that for suitable choice of the parameters $ k_{nR}$ and
$ k_{nI}$, the solitary waves get peaked around two values as shown
in Figs. 2-3.

Now to investigate the effect of non-zero values of $ \rho_1, \rho_2, \tau_1 $ and $ \tau_2 $
or in other words to see the nontrivial contributions due to cubic-quintic
generalization   (\ref{new}), we evaluate    
$ \ 
q_{jt} = \left[ q_{jMt} + i q_{jM} \theta_{jt} \right] exp(i \theta_{j})
,\qquad j=1,2. 
\ $ and plot the asymptotic behavior of the
 two-soliton solution (19) to the quintic generalization of the Manakov
model Eq.(\re{new}) in Figs. 4a,b. The corresponding interaction profile
of the solitons during their propagation is shown in Fig. 5.
 We  observe firstly that,  
 as in the case of Manakov model, here also generically 
the fascinating shape changing inelastic collision persists. However,
in this case  
 one can overcome the splitting
effect of Figs. 2-3 corresponding to the Manakov model.
  For example, if we set $ \rho_1 = \rho_2 = \tau_1 = \tau_2 = 1 $,
in (19) then the splitting of solitons   disappear, as evident from Figs. 4-5.
The reason for this is that now we have
$$ 
|q_{jt}|^2 =  |q_{jMt}|^2 + | q_{jM}|^2 | \theta_{jt}|^2=
 |q_{jMt}|^2 + | q_{jM}|^2 (\de_1 | q_{1M}|^2 +\de_2 | q_{2M}|^2), 
$$
where $\de_1$ and $\de_2$ are equal to $\rho_1$ and $\rho_2$ for $j=1$,
while to $\tau_1$ and  $\tau_2$ for $j=2$. Since here
 the second term dominates over the first in the region of splitting, the 
splitting   effect naturally gets supressed.
 Comparing  Figs. 3 and 5 it is
also important to note  that the
 intensity of solitons ( $ |q_{1t}|^2,
|q_{2t}|^2 ) $ at the intersection region for  
solution (19)  of our generalized model is much higher than
that for (20) corresponding to  the Manakov model.
  The above processes vividly demonstrate
 the nontrivial effect of the additional terms involving
parameters
 $ \rho_1$, $\rho_2$, $\tau_1$ and $ \tau_2$ appearing in Eq.(\ref{new}).
 
\section {Relation with the Manakov model}

As we have mentioned above there exists an interesting interrelation  between the
quintic generalization (\re {new}) and the Manakov model (\ref{manakov}),
which in fact we have used already in deriving the soliton solutions  of 
(\re {new}). We establish now this relationship by showing that the Lax
operators of these two models are   related through a local gauge
transformation \c{29,30,31},
 while the fields are connected by a nonlinear transformation
 in dependent variables.

It is known \c{32} that under a gauge transformation of the Jost function
$\Phi^{'}=g
\Phi$ with the gauge field $g \in U(3)$,  the Lax operators transform  as 
\be
L^{\prime } =g^{-1}Lg-g^{-1}g_t,  \ \quad
M^{\prime } =g^{-1}Mg-g^{-1}g_z.  \label{gt}
\ee
Choosing now  the specific 
form
\begin{equation}
g= \left(
\begin{array}{ccc}
1 & 0 & 0 \\ 
0 & \exp (-i\theta_1) & 0 \\ 
0 & 0 & \exp (-i\theta_2) 
\end{array}
\right),  \label{g}
\end{equation}
 with its
elements $\ \th_1, \th_2$ being the same functions of z and t as in Eq. (11c)
and performing the transformation (\re{gt}), one can conveniently remove
the diagonal  terms involving $\th_{1t},\th_{2t}$ and $\th_{1z},\th_{2z}$
 in the Lax pair (11a,b). It can be observed
further that the resultant gauge transformed Lax operators reduce exactly
to those of the Manakov model \c{22,33} if we introduce  transformed fields 
 \be q_{aM}= q_a\exp (-i\theta _a), \quad  a=1,2
\ll{q-q}\ee
 along with their conjugates. At the same time transformation (\re{q-q})
reduces the equations (\re{new}) to those of the Manakov model
(\re{manakov}).

The above points establish the relationship between these models and
justifies the form of soliton solution presented in the earlier section for
the model (\ref{new}). Moreover, this procedure also provides an alternative proof of
the integrability of our model.  It is important to note that under such
gauge transformation the Poisson bracket structure of the fields also gets
changed. To find such changes in the canonical structure we may use
transformation (\re{q-q}) to express our field through the Manakov fields
and assuming
 standard canonical relation  (\re{can}) for the Manakov model, we can
 derive the anyon-like relations for the 
fields of (\re{new}):
 \bea \{q_1(x), q^*_1  (y)\}&=& \de(x-y)+
i \rho_1 \epsilon (x-y)   q_1(x) q^*_1 (y) ,
 \nonumber \\  \{q_1(x), q_1 (y)\} &=&
 i \rho_1 \epsilon (y-x)   q_1(x) q_1 (y) ,
\nonumber
\\  \{q_1(x), q_2 (y)\}&= &- i (\rho_2 \theta (x-y)- \tau_1 \theta (y-x))
 q_1(x) q_2 (y),
\nonumber
\\  \{q_1(x), q_2^* (y)\}&= &i (\rho_2 \theta (x-y)- \tau_1 \theta (y-x))
 q_1(x) q_2^* (y),
 \ll{noncan}
 \eea
etc., where $\epsilon (x) =\theta (x)-\theta (-x)$ is the sign function
defined through the step function: $\theta (x)= 1 $ for $ x>0, \ \theta
(x)=0,$ for $x \le 0$.
 Note that at $x=y$ the fields exhibit canonical property, while at $x
\neq y$ their behavior is nonultralocal and mimics anyon-like properties
\cite {kundu98} in the classical limit.
It may be remarked 
 here that the generalized Manakov equation (9) can be derived
directly from the Hamiltonian (17) by  careful application of the PB
structure (25) and the relation $\partial_x \theta (x-y) =\de (x-y)$.
\section{Vector generalization of chiral solitonic model}
We have seen that for obtaining (\re{new}) from the Hamiltonian (\re{c3}) we have to
use  noncanonical brackets (\re{noncan}).
 On the other hand if nevertheless one
considers them to be canonical, i.e. \be \{q_i(x), q^*_j (y)\}=\de(x-y)
\de_{ij}, \ \{q_i(x), q_j (y)\}=0, \ll{can} \ee
 from the same Hamiltonian (\re{c3}) we can derive  completely  different
coupled equations with
fifth-degree nonlinearity. If for simplicity we assume
$\rho_1=\rho_2=\tau_1=\tau_2 =\rho_0$, we can
 derive these equations easily from 
 $c_3$ as
\be
iq_{1z}+q_{1tt}+ 2 (|q_1|^2 +
|q_2|^2)q_1-3 \rho_0 ^2(|q_1|^2 +|q_2|^2)^2q_1-2i \rho_0 (
 (|q_1|^2+|q_2|^2)q_{1t}+(q_{1}^*q_{1t}+q_{2}^*q_{2t})q_1) 
= 0 \ll{new1} \ee
and similarly for $q_2$ by interchanging the indices $1 \leftrightarrow 2$
in (\re{new1}).
We  notice that this system of coupled equations again with cubic-quintic
nonlinearity is a new system  which is different from Eq. (\re{eqn})
 presented earlier. To anlyse these equations
more closely we perform again a nonlinear variable change as 
$q_a \rw Q_a=q_a e^{-i \rho_0 \th}$ with $\th_t=N \equiv |Q_1|^2+|Q_2|^2$.
After some lenghty but simple manipulations one can reduce the system
(\re{new1}) further to a more compact form with only cubic nonlinearity:
\be 
iQ_{az}+Q_{att}+2(N-\rho_0 j) Q_a=0,
\ll{new2}
\ee
where we have denoted $j=j_1+j_2, j_a= i(Q^*_aQ_{at}-Q_aQ^*_{at}) .$
We immediately  recognize that this is nothing but the  vector generalization
of the Aglietti  et al equation \c{jakiw},  however with the addition of
 a cubic
nonlinearity  coming from the Manakov model.
Nevertheless the system (\re{new2}) shows remarkable property close to the
chiral-soliton feature of \c{jakiw}. In particular assuming the 1-soliton
form as $Q_a=A_a s(t-vz) e^{i({ v \ov 2} t+ \omega z)}$ one may conclude
that here the quantity $\kappa =1+ v \rho_0$ acts as the effective coupling
constant of the nonlinear term, which regulates the intensity of the soliton.
Therefore for the soliton velocity  $v > -{1 \ov \rho_0} $ 
only (with $\rho_0>0$) Manakov model
like 
bright solitary wave solution can exist. With decreasing velocity the effective coupling
constant $\kappa$ also decreases, which interestingly 
  influences the intensity of the soliton
to increase and
reflects a possible nonintegrable property of the model.  Finally for the
soliton velocity 
$v = -{1 \ov \rho_0} $ the nonlinear term  to sustain the soliton disappears
and hence no such soliton can appear anymore. However for the negative
velocity below this value, $v <  -{1 \ov \rho_0} ,$ the sign of $\kappa$ flips
and kink-like exact dark soliton can appear. For $\rho_0<0$ the whole
picture reverts. This amazing solitonic feature evidently is a
generalization of the chiral soliton property  of \c{jakiw} due to the
presence of Manakov term as well as the multi-component nature and may have
important  applications in nonlinear optics. The suspected nonintegrable
nature of this system and consequently the original chiral-solitonic system
\c{jakiw} can be convincingly proved by showing that the conserved
quantities  of the model are not in involution (in particular using the 
canonical bracket (\re{can}) it can be shown that $ \{c_2,c_3\} \not = 0$).
Therefore though this system possesses  Lax pair and infinite conserved
quantities, their noninvolutiveness spoils the integrability. The involution
of the conserved quantities however is restored if we use the noncanonical 
bracket (\re{noncan}) and this ensures the exact integrability of (\re{new}). 
\section{Conclusion}
We have constructed the  Lax pair of the  
proposed integrable CNLS equation (\ref{new}) with cubic-quintic nonlinearity 
governing the soliton propagation in non-Kerr media,
and using it generated the infinite set of its conserved quantities in
 explicit form. We also presented the exact one and two-soliton solutions
of the model using those of the well known 
 Manakov model. It has been demonstrated through   the explicit two-soliton
solution of the proposed model 
that    the intensity of the 
t-derivative of  soliton in the interaction region
 is  much higher than that of the Manakov model. Moreover,
 the localized part of the time derivative of the Manakov soliton  gets
splitted and
peaks around two values as shown
in Figs 2 and 3. However, such splitting
can be suppressed  in the generalized cubic-quintic equation (\ref{new})
having  nonzero $ \rho $'s and $ \tau$'s as 
 has been demonstrated    in Figs. 4 and 5. These figures also confirm that
  the shape changing
inelastic soliton collisions, as in the Manakov case,  persist in our
model. 
We believe that  our results will   be  found equally  useful  in  more
general 
situations like Eq. (5) by taking our model   as the unperturbed part
  and treating the remaining terms as perturbations.

We have also established the relationship between the proposed model and the Manakov
model at the Lax pair level as well as at the field solution level, which
 shows  an intriguing change in the
canonical structure, namely the bosonic relations of Manakov model
transforms into  the anyonic relations of the present system.

Another remarkable fact is that   
assuming the standard canonical structure for our fields we are able to
derive from the same Hamiltonian yet another
  coupled system with cubic-quintic nonlinearity. This novel
model, which turns out to be nonintegrable, represents
a vector generalization of Aglietti et al model, famous
for exhibiting chiral-soliton solutions. Such
chiral-soliton property also prevails in the present vector case showing
fascinating properties of the solitons, like changing intensity with soliton
velocity, vanishing of bright solitons and
appearance of dark solitons   below certain velocity etc. Such properties   may have
important applications in nonlinear optical processes.  

\section{Acknowledgements}
The work of R.R and M.L. forms part of a Department of Science and Technology
research project.

\section*{Figures}

\begin{figure}
\centerline{\epsfig{figure=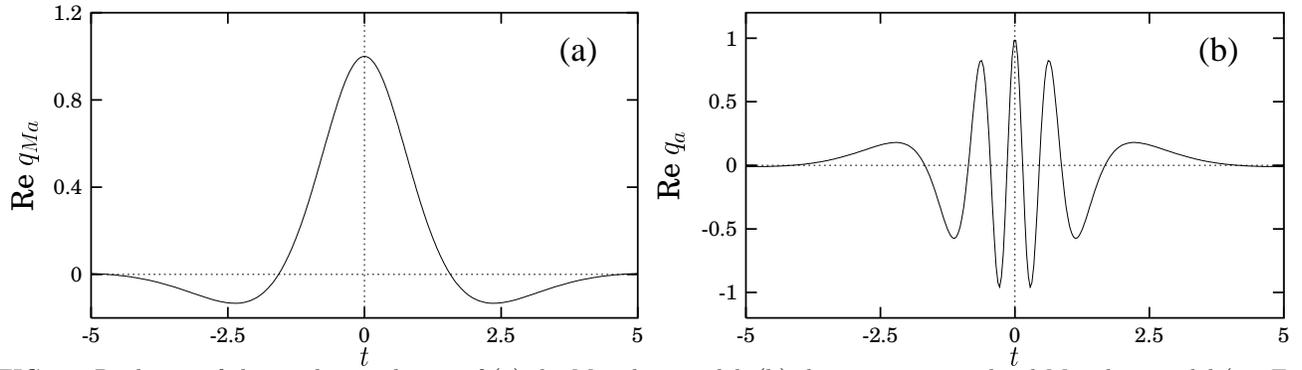, width=\linewidth}}
\caption{ Real part of the 1-soliton solution of (a)  the Manakov
model,  (b) the quintic generalized Manakov model (see Eqn.
(\re{s1})).  
}
\label{fig1}
\end{figure}

\begin{figure}
\centerline{\epsfig{figure=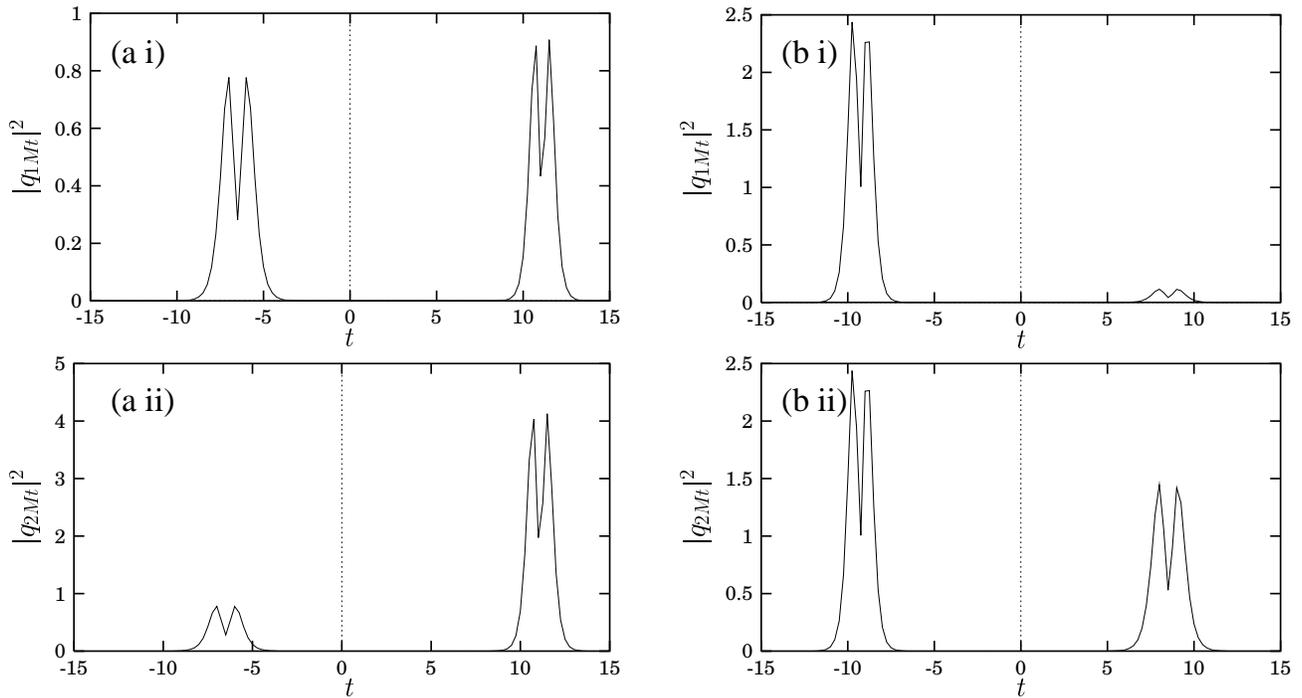, width=\linewidth}}
\caption{Asymptotic forms of the intensity profiles $ |q_{1Mt}|^2 $ and $
|q_{2Mt}|^2 $ of the two-soliton solution (20) of the Manakov model ( $
\rho_1 = \rho_2 = \tau_1 = \tau_2 = 0 $)  with the parameter values $
k_1 = 1.5 + i 0.5 $, $ k_2 = 2.0 - i 0.7 $, $ \alpha_1 =  \beta_1 =
\beta_2 = 1 $, $ \alpha_2  = {(39 + i 80) \over 89} $ , (a) at $ z = -7
$ (before interaction) , (b) at $ z = +7 $ (after interaction). Note
the splitting in the asymptotic soliton profiles.
}
\label{fig2}
\end{figure}

\begin{figure}
\centerline{\epsfig{figure=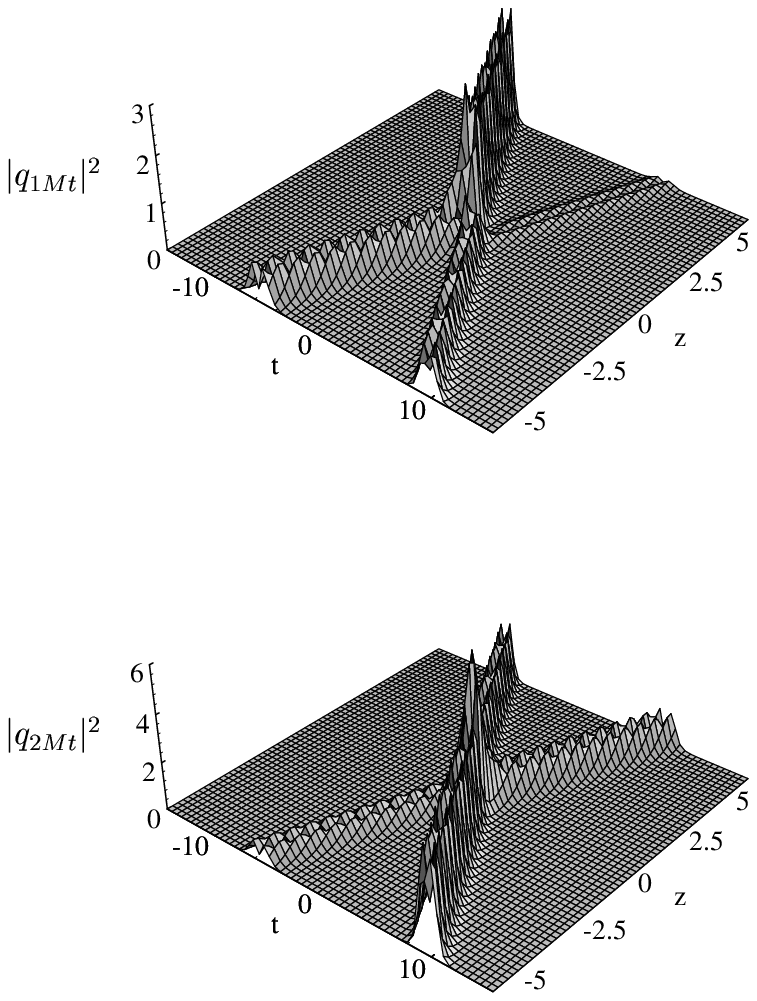, width=.5\linewidth}}
\caption{Intensity profiles $ |q_{1Mt}|^2 $ and $ |q_{2Mt}|^2 $ of the
two-soliton solution of the Manakov model with the parameteric values
as in Fig.~\ref{fig2}
}
\label{fig3}
\end{figure}

\begin{figure}
\centerline{\epsfig{figure=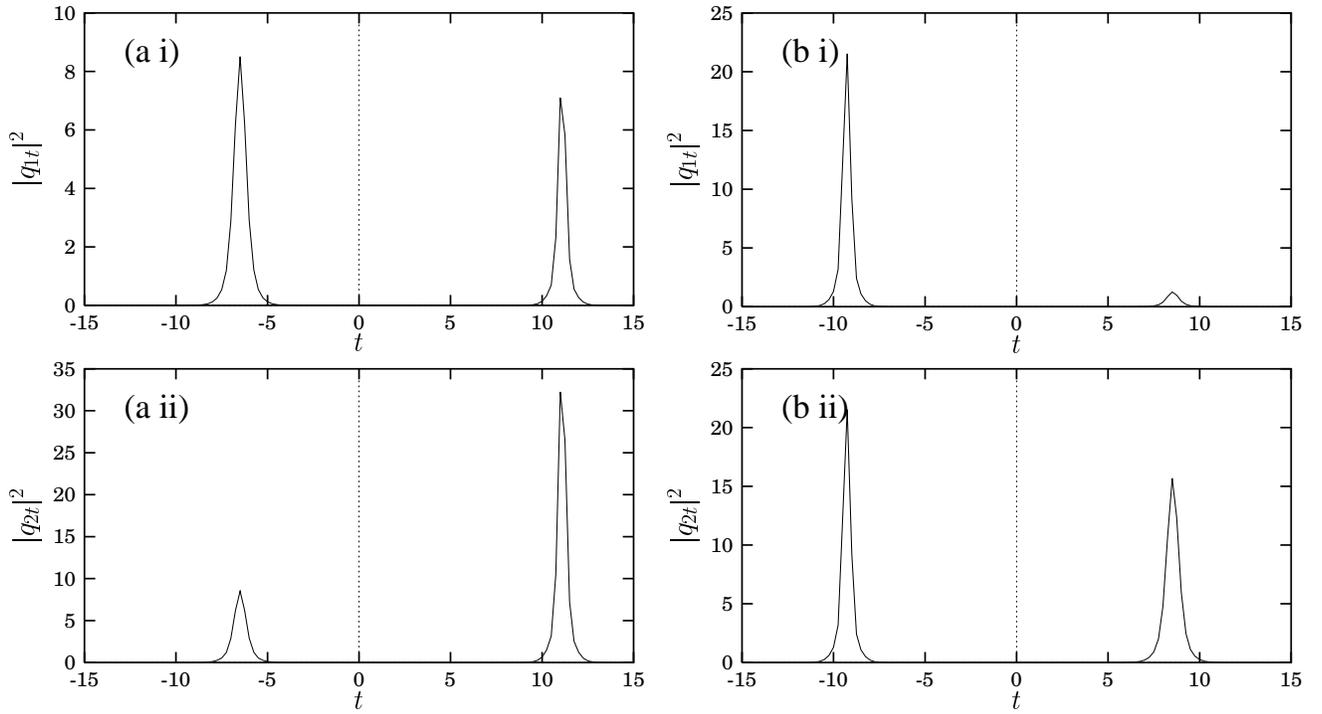, width=\linewidth}}
\caption{Asymptotic forms of the intensity profiles $ |q_{1t}|^2 $ and
$ |q_{2t}|^2 $ of the two-soliton solution (19) of  the generalized
model (\re{new}),  (a) at $ z=-7 $, (b) $ z = +7 $ , for nonzero values
of the parameters $ \rho_1 = \rho_2 = \tau_1 = \tau_2 = 1 $ and with
remaining parameters  as in Fig. 3.  Note the suppression of the
soliton splitting, which appeared in the asymptotic profile in Figs.~\ref{fig2}
and \ref{fig3}.
}
\label{fig4}
\end{figure}

\begin{figure}
\centerline{\epsfig{figure=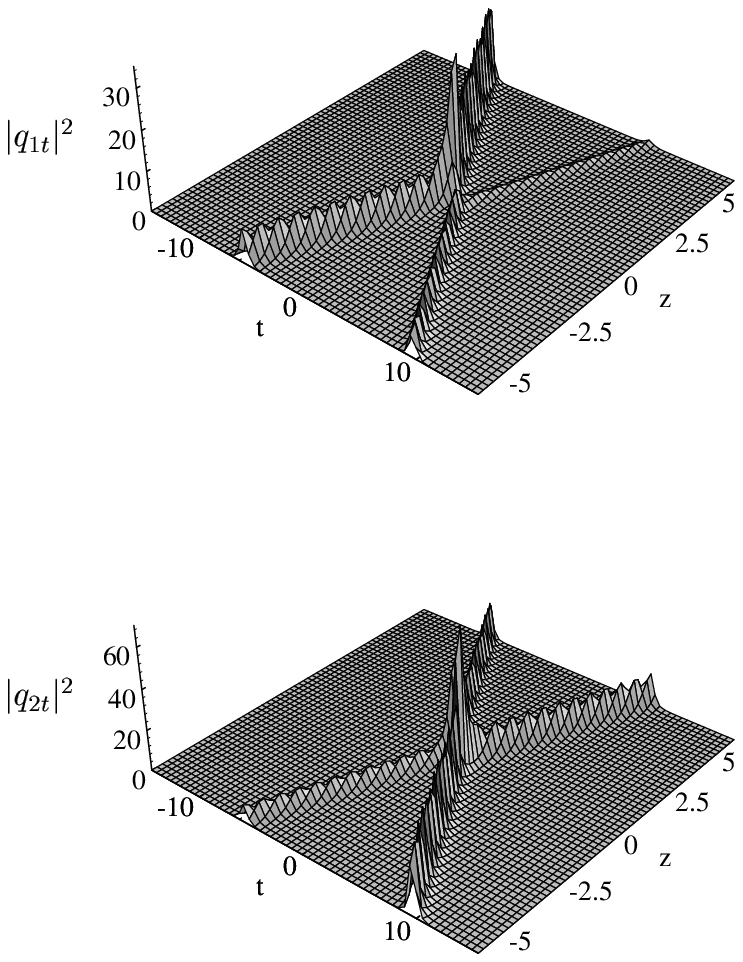, width=.5\linewidth}}
\caption{Intensity profile $ |q_{1t}|^2 $ and $ |q_{2t}|^2 $ of the
two-soliton solution (19) of the generalied Manakov model with the
parameter values as in Fig. 4.  Note the persistence of inelastic
soliton collision as in the Manakov model and a higher intensity of
modes during soliton interaction compared  to the Manakov model.
}
\label{fig5}
\end{figure}


\references
\bibitem{1}
G.P.Agarwal, {\it Nonlinear Fiber Optics} - Second Edition (Academic
Pres, New York, 1995).
\bibitem{2}
Yu. S. Kivshar and B. L. Davies, Phys. Rep. {\bf 298}, 81 (1998).
\bibitem{3}
N. Akhmediev and A. Ankiewicz, {\it Solitons: Nonlinear Pulses and Beams}
(Chapman $ \& $ Hall, London, 1997).
\bibitem{4}
H. A. Haus and W. S. Wong, Rev. Mod. Phys. { \bf 68}, 423 (1996).
\bibitem{5}
A. Hasegawa and Y. Kodama, {\it Solitons in Optical Communications}
(Oxford University Press, England, 1995).
\bibitem{6}
M. N. Islam, {\it Ultrafast Fiber Switching Devices and Systems} (Cambridge
University Press, England, 1992).
\bibitem{7}
A. C. Newell and J. V. Moloney, {\it Nonlinear Optics} (Addison-Wesley, New York,
1992).
\bibitem{8}
F. Abdullaev, S. Darmanyan, P. Khabibullaev, {\it Optical Solitons}
(Springer-Verlag, Berlin, 1993).
\bibitem{9}
A. E. Kaplan, Phys. Rev. Lett. {\bf 55}, 1291 (1985);
R. H. Enns, S. S. Rangenekar and A. E. Kaplan, Phys. Rev. {\bf A36},
1270 (1987);
R. H. Enns, D. E. Edmundson, Phys. Rev. {\bf A47}, 4254 (1993);
A. Kumar, T. Kurz and W. Lauterbon, Phys. Lett {\bf A235}, 367 (1997).
\bibitem{10}
D. I. Pushkarov and S. Tanev, Opt. Commun. {\bf 124}, 354 (1996):
S. Tanev and D. I. Pushkarov, Opt. Commun. {\bf 141}, 322 (1997).
\bibitem{11}
A. Kumar and A. Kumar, Opt. Commun. {\bf 125}, 377 (1996).
\bibitem{12}
J.M. Soto-Crespo and L. Pesquera, Phys. Rev.  {\bf E56}, 7288 (1997).
\bibitem{13}
C. Zhou, X. T. He and S. Chen, Phys. Rev. {\bf A46}, 2277 (1992).
\bibitem{14}
V. V. Afanasjev, P. L. Chu and Yu. S. Kivshar, Opt. Lett.
{\bf 22}, 1388 (1997).
\bibitem{15}
V. Skarka, V. I. Berezhiani and R. Miklaszewski, Phys. Rev. {\bf E56},
1080 (1997).
\bibitem{16}
D. Artigas, L. Torner, J. P. Torres and N. Akhmediev, Opt. Commun.
{\bf 143}, 322 (1997); P. Honzatko, Opt. Commun. {\bf 127}, 363 (1996).
\bibitem{17}
G. Dattoli, F. P. Orisitto and A. Toree, Opt. Lett. {\bf 14}, 456 (1989).
\bibitem{18}
A. W. Snyder and A. P. Sheppard, Opt. Lett. {\bf 18}, 482 (1993).
\bibitem{19}
C. Pare and M. Florjanczyk, Phys. Rev. {\bf A41}, 6287 (1990) and
references therein.
\bibitem{20}
M. Hisakado, T. Iizuka and M. Wadati. J. Phys. Soc. Jpn, {\bf 63}, 2887 (1994).
\bibitem{21}
M. Hisakado and M. Wadati, J. Phys. Soc. Jpn, {\bf 63},  3962 (1994); {\bf 64},
408 (1995).
\bibitem{22}
S. V. Manakov, Zh. Eksp. Theor. Fiz. {\bf 65}, 505 (1973)
[Sov. Phys. JETP {\bf 38}, 248 (1974)].
\bibitem{23}
R. Radhakrishnan, M. Lakshmanan and J. Hietarinta, Phys. Rev. {\bf E56},
2214 (1997).
\bibitem{24}
A. Hasegawa and T. Nyu, J. Lightwave Technol. {\bf LT11}, 395 (1993).
\bibitem{25}
M. H. Jakubowski, K. Steiglitz and R. Squiev, Phys. Rev. {\bf E58}, 6752 (1998).
\bibitem{26}
A. Ankiewicz and N. Akhmediev, Opt. Commun. {\bf 124}, 95 (1996).
\bibitem{27}
S. Wabnitz, S. Trillo, E. M. Wright and G. I. Stegeman, J. Opt. Soc. Am.
{\bf B8}, 602 (1991); S. Trillo, S. Wabnitz, W. C. Banyai, N. Finlayson,
C. T. Seaton, G. I. Stegeman and R. H. Stolen, IEEE. J. Quantum Electron
{\bf QE25}, 104 (1989).
\bibitem{28}
R. Radhakrishnan and M. Lakshmanan, Phys. Rev. {\bf E54}, 2949 (1996).
\bibitem{29}
A. Kundu,  J. Math. Phys, {\bf 25}, 3433 (1984)
;
F. Calogero, Inverse Problem, {\bf 3}, 229 (1987)
;
Li Shen in {\it Symmetries \& singularity structures} (Springer, Ed. M.
Lakshmanan, 1990), 27
\bibitem{30}
S. Kakei, N. Sasa and J. Satsuma, J. Phys. Soc. Jpn, {\bf 64}, 1519 (1995).
\bibitem{31}
K. Kondo, K. Kajiwara and K. Matsui, J. Phys. Soc. Jpn, {\bf 66}, 60 (1997).
\bibitem{32}
V. I. Zakharov and L. A. Takhtajan, Teor. Mat. Fiz. {\bf 38}, 26 (1979); 
M. Lakshmanan (Ed.), {\it Solitons: Introduction and Applications} (Springer-
Verlag, Berlin, 1988), p. 86.
\bibitem{jakiw} U. Aglietti, L. Griguolo, R. Jackiw, S.Y. Pi and D.
Seminara, Phys. Rev. Lett. {\bf 77}, 4406 (1996)
\bibitem{kundu98} A. Kundu, e-print hep-th/9811247 {\it Exact solution of double-$\delta$ function Bose gas through
interacting anyon gas}. 

\bibitem{33}
D. J. Kaup and B. A. Malomed, Phys. Rev. {\bf A48}, 599 (1993);
M. Karlsson, D. J. Kaup and B. A. Malomed, Phys. Rev. {\bf E54}, 5802 (1996).
\bibitem{34}
R. Radhakrishnan and  M. Lakshmanan  J. Phys. A: Math. Gen {\bf 28}, 2683 (1995).
\end{document}